\documentclass[12pt]{article}
\usepackage[scaled]{helvet}
\usepackage[margin=1in]{geometry}
\RequirePackage[OT1]{fontenc}
\RequirePackage{amsthm,amsmath}

\usepackage{amsmath}
\usepackage{graphicx,psfrag,epsf,subfig}
\usepackage{enumerate}
\usepackage{natbib}
\usepackage[hidelinks]{hyperref}
\usepackage[graphicx]{realboxes}

\usepackage[nomarkers]{endfloat}

\usepackage{multicol, booktabs, subfig, bbm, color, xcolor}
\usepackage{amssymb}

\usepackage[margin=1in]{geometry}
\usepackage{setspace}

\usepackage{titlesec}

\titleformat*{\section}{\normalsize\bfseries}
\titleformat*{\subsection}{\normalsize\it}
\usepackage[affil-it]{authblk}

\graphicspath{{img/}}

\title{Bayesian Nonparametric Monotone Regression}

\author[1]{Ander Wilson\footnote{ander.wilson@colostate.edu}}
\author[2]{Jessica Tryner}
\author[2]{Christian L'Orange}
\author[2]{John Volckens}
\affil[1]{Department of Statistics, Colorado State University}
\affil[2]{Department of Mechanical Engineering, Colorado State University}
\date{}

\begin{document}

\maketitle

\setstretch{1.7}

\begin{abstract}
\singlespacing 
In many applications there is interest in estimating the relation between a predictor and an outcome when the relation is known to be monotone or otherwise constrained due to the physical processes involved. We consider one such application--inferring time-resolved aerosol concentration from a low-cost differential pressure sensor. The objective is to estimate a monotone function and make inference on the scaled first derivative of the function. We proposed Bayesian nonparametric monotone regression which uses a Bernstein polynomial basis to construct the regression function and puts a Dirichlet process prior on the regression coefficients. The base measure of the Dirichlet process is a finite mixture of a mass point at zero and a truncated normal. This construction imposes monotonicity while clustering the basis functions. Clustering the basis functions reduces the parameter space and allows the estimated regression function to be linear. With the proposed approach we can make closed-formed inference on the derivative of the estimated function including full quantification of uncertainty. In a simulation study the proposed method performs similar to other monotone regression approaches when the true function is wavy but performs better when the true function is linear. We apply the method to estimate time-resolved aerosol concentration with a newly-developed portable aerosol monitor. The {\tt R} package {\tt bnmr} is made available to implement the method.

\vspace{1em}

\noindent Keywords: Bernstein polynomials; Dirichlet process; monotone regression; Aerosol monitors; Fine particulate matter
\end{abstract}

\newpage

\section{Introduction}

In environmental, biomedical, and engineering applications a common objective is to estimate the relation between a predictor and an outcome when there is prior knowledge that the relation is monotone or otherwise shape-constrained. In this paper we consider one such application that relates to measuring airborne particles at fine-temporal resolution using a recently-developed portable monitor. At the center of this problem is estimation of a function that is known  to be monotone due to the physical processes involved in the monitor and making inference on the scaled first derivative of the estimated monotone function which is equal to estimated aerosol concentration. 

Measuring air pollution with high temporal and spatial resolution is critical to both conducting air pollution research and protecting the public's health. In an ideal world, we would be able to use a large number of monitors to measure personal air pollution exposure in cohort studies of health effects or to deploy in networks to warn of potential risks such as those from exposure to wildfire smoke. However, the large size and high cost of air quality monitors has historically prohibited widespread use. Hence, there is a need to develop smaller, more affordable monitors and the accompanying data science tools to make meaningful inference on the readouts of these monitors. 

In this paper we consider inference for data generated by the recently-developed Mobile Aerosol Reference Sampler (MARS). MARS was designed to be an affordable, portable monitor for measuring fine particulate matter ($\text{PM}_{2.5}$) concentrations in environmental and occupational health studies \citep{Tryner2019a}. The MARS device is built on the Ultrasonic Personal Aerosol Sampler (UPAS) platform which has also been previously described in the literature \citep{Volckens2017}. MARS uses a piezoelectric microblower to pull air through a PM$_{2.5}$ cyclone inlet and a 25mm filter. A high-resolution pressure sensor measures the time-resolved pressure drop across the sampling filter. As particles accumulate on the filter the pressure drop across the filter increases. This pressure drop should be positive and increase monotonically in time during measurement. Deviations from monotonicity only occur (1) in the first few minutes of use when a new filter is stretching out or (2) if there is a change in air density or particle source. In the experimental data used in this paper, particle source remained constant and only minor changes in air density occurred. Time-resolved PM$_{2.5}$ concentration can be inferred from the time-resolved rate of change in pressure drop after the latter is normalized to the total PM$_{2.5}$ mass collected on the filter. Specifically, when the derivative is scaled so that the area under the derivative function is equal to total PM$_{2.5}$ mass collected on the filter divided by volumetric flow rate, then the scaled first derivative is a measure of PM$_{2.5}$ concentration as a function of time \citep{Novick1992,Dobroski1997}. Hence, the objective is to estimate the pressure drop as a function of time and then make inference on the scaled first derivative of pressure drop.

Several approaches have been proposed to estimate monotone functions. Early works include estimation of shape-constrained piecewise linear functions \citep{Hildreth1954, Brunk1955}. \cite{Mammen1991} proposed monotone kernel smoother methods and \cite{Mammen2001} proposed monotone projections of unconstrained smooth estimators. A large number of spline based approaches have been proposed including cubic smoothing splines \citep{Wang2008a}, constrained regression splines \citep{Ramsay1988,Meyer2008,Meyer2011, Powell2012}, penalized splines \citep{Meyer2012}, and piecewise linear splines \citep{Neelon2004}. Several recent papers have proposed monotone Bernstein polynomial (BP) regression \citep{Chang2005,Chang2007,Curtis2011, Wang2011,Wilson2014c,Ding2016}. 

In this paper we take a BP approach to constrained regression.  Monotonicity can be imposed with BPs by imposing a linear order constraint on the regression coefficients. An alternative but equivalent approach is to linearly transform the regression coefficients and then impose a positivity constraint on all of the transformed regression coefficients with the exception of the intercept, which is unconstrained \citep{Wang2011}.  \cite{Curtis2011} proposed a variable selection approach to monotone regression with BPs that puts a variable selection prior on the transformed regression coefficients akin to a mixture of a mass point at 0 and a normal distribution truncated below at 0. The approach is appealing because it imposes monotonicity, allows for data-driven tuning of the model by selecting excess basis functions out of the model and allows for no association when all coefficients are selected out of the model. 

The approach we present here, which we refer to as Bayesian nonparametric monotone regression (BNMR), is similar to that of \cite{Curtis2011} in that we use a BP expansion and  a variable selection prior that imposes monotonicity. In contrast, our approach both selects some regression coefficients to be zero and clusters other regression coefficients. By clustering regression coefficients we create a reduced set of combination basis functions that are  each the sum of multiple BPs and assigned a single regression coefficient. This has two distinct advantages over only variable selection. First, when all regression coefficients are clustered together into a single combination basis function the approach is equivalent to performing linear regression with the slope constrained to be non-negative. This improves performance when the true regression function is in fact linear. Second, when the true regression function is nonlinear our approach requires a reduced number non-zero regression coefficients each corresponding to the combination of a mutually exclusive set of basis functions. In a simulation study we show that our approach is able to match the flexibility of alternative approaches but uses a smaller number of parameters. As a result our Markov chain Monte Carlo (MCMC) approach samples from the full conditional of a truncated multivariate normal distribution of smaller dimension which can reduce autocorrelation in the resulting chain. Hence, the proposed method allows for flexible monotone regression while allowing the model to be null when there is no association between predictor and outcome and allowing the function to be linear when there is no evidence of nonlinearity. This results in comparable performance to other approaches for smooth nonlinear functions but improved inference when the true relation is linear.

We apply the proposed approach to evaluate 12 samples collected using MARS in a controlled laboratory chamber. We compare estimated time-resolved PM$_{2.5}$ inferred with the proposed method based on 30-second measurements of pressure drop across the MARS filter to minute-resolution measurements of PM$_{2.5}$ in the chamber reported by a tapered element oscillating microbalance (TEOM) (1405 TEOM, ThermoFisher Scientific, Waltham, MA, USA), which is a regulatory-grade PM$_{2.5}$ monitor.

\section{Methods}

\subsection{Model formulation}\label{sub:model}

Our primary interest is estimating the regression function
\begin{equation}
\label{eq:model}
y_i=f(x_i) + \epsilon_i
\end{equation}
where $f$ is an unknown monotone function. Without loss of generality, and consistent with our application, we assume that $f$ is monotone increasing. We also assume that $x$ is scaled to the unit interval.

\begin{figure}
    \centering
    \subfloat[Bernstein polynomial basis]{
    \includegraphics[]{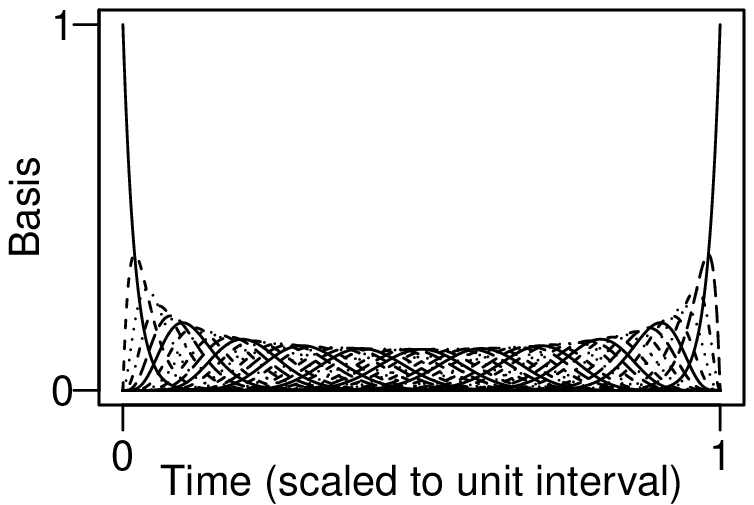}
    \label{subfig:BP}
    }
    \subfloat[Transformed Bernstein polynomial basis]{
    \includegraphics[]{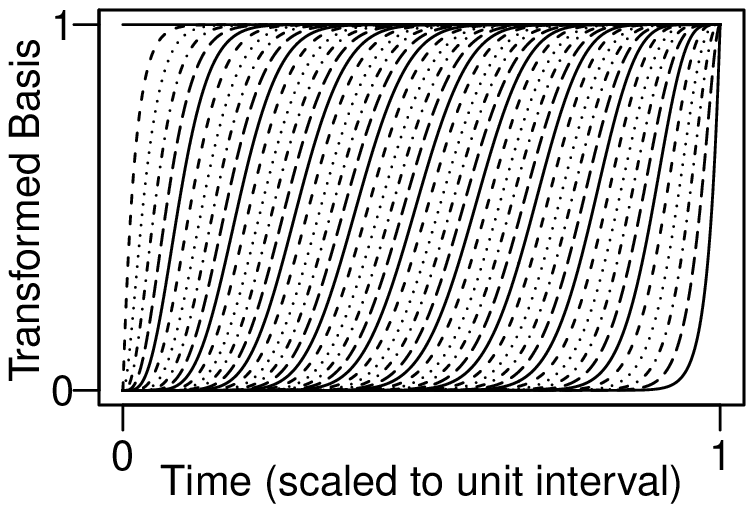}
    \label{subfig:transBP}
    }

    \subfloat[Selected transformed Bernstein polynomial basis used with BISOREG]{
    \includegraphics[]{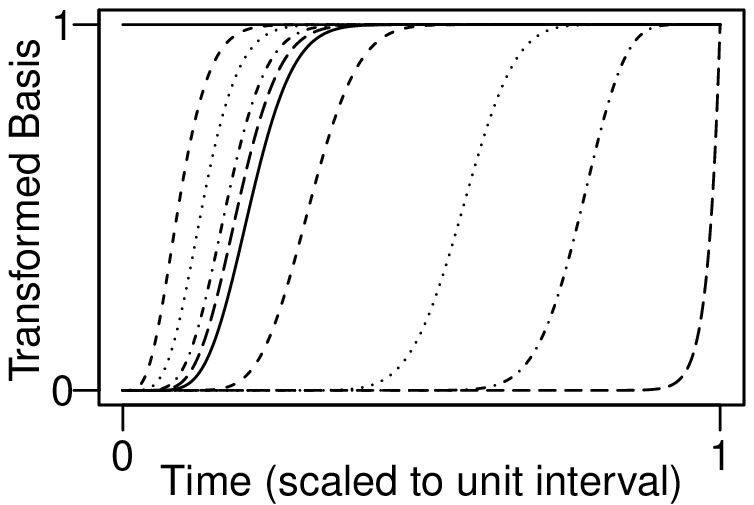}
    \label{subfig:bisoregBP}
    }
    \subfloat[Linear combination of transformed Bernstein polynomial basis used with BNMR]{
    \includegraphics[]{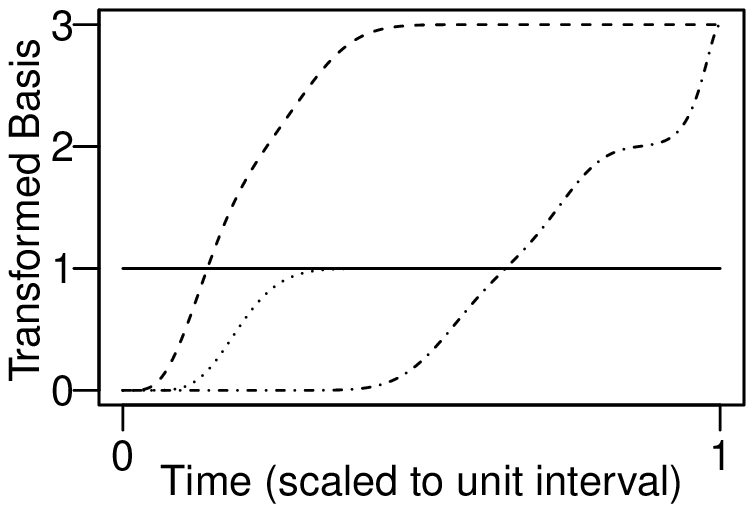}
    \label{subfig:bnmrBP}
    }
    
    \caption{Various representations of the Bernstein polynomial (BP) basis functions. Panel~\ref{subfig:BP} shows the 51 BP basis functions of order $M=50$ ($\Psi(x,M)$). Panel~\ref{subfig:transBP} shows the transformed BP basis represented as $\Psi(x,M) \mathbf{A}^{-1}$ as described in ~\ref{sub:model}. This transformation is used for both BNMR and BISOREG.   Panel~\ref{subfig:bisoregBP} shows the posterior mode group of basis functions selected to be included into the model with BISOREG. This is a subset of the transformed basis functions shown in Panel~\ref{subfig:BP}. Panel~\ref{subfig:bnmrBP} shows the posterior model combination of basis functions included with BNMR. This includes the intercept and three basis functions which are each a linear combination of one to three of the basis functions shown in panel~\ref{subfig:transBP} and subsequently linear combinations of the basis functions shown in Panel~\ref{subfig:BP}. Results from all 12 runs are shown in the supplemental material.}
    \label{fig:BP}
\end{figure}

We parameterized $f$ using a BP expansion. The $k^\text{th}$ BP basis function of order $M$ is
\begin{equation}
\label{eq:bpbasis}
\psi_k(x,M) = \left(\begin{array}{c}M \\ k \end{array}\right) x^k (1-x)^{M-k}.
\end{equation}
The regression function expressed as a weighted combination of BPs is 
\begin{equation}
f(x) = \sum_{k=0}^M \psi_k(x,M)\beta_k = \Psi(x,M)\boldsymbol\beta,
\label{eq:fexpand}
\end{equation}
where $\boldsymbol\beta=\left(\beta_0,\dots,\beta_M\right)^T$ are regression coefficients and $\Psi(x,M)=\left[ \psi_0(x,M),\dots,\psi_M(x,M)\right]$. The first regression coefficient $\beta_0$ parameterizes the intercept. Figure~\ref{subfig:BP} shows the BP basis used in the data analysis.

The regression function in \eqref{eq:fexpand} is monotone increasing if $\beta_{k-1}\le\beta_k$ for all $k=1,\dots,M$. Following \cite{Curtis2011}, it is convenient to reparameterize the regression coefficients.  Let $\mathbf{A}\boldsymbol\beta=\boldsymbol\theta$ where $\boldsymbol\theta=\left(\theta_0,\dots,\theta_M\right)^T$ and the $(M+1)\times (M+1)$-matrix $\mathbf{A}$ is such that $\theta_0=\beta_0$ and $\theta_k=\beta_k-\beta_{k-1}$ for $k=1,\dots,M$:
\begin{equation}
\label{eq:Amonotone}
\mathbf{A} = \left[\begin{array}{cccccc}
 1 &  0 &  0 &  \dots &  0 &  0 \\
-1 &  1 &  0 &  \dots &  0 &  0 \\
 0 & -1 &  1 &  \dots &  0 &  0 \\
\vdots & \vdots & \vdots & \ddots & \vdots & \vdots \\
 0 & 0 &  0 &  \dots &  -1 &  1 \\
\end{array}\right].
\end{equation}
The regression function is then 
\begin{equation}
\label{eq:reparam}
f(x) = \Psi(x,M) \mathbf{A}^{-1} \boldsymbol\theta.
\end{equation}
Figure~\ref{subfig:transBP} shows the transformed basis $\Psi(x,M) \mathbf{A}^{-1}$ used in the data analysis.

Using this reparameterization $f$ is monotone increasing when $\theta_k\ge0$ for all $k>0$. Further, $f$ is linear with the form $f(x)=\theta_0+wx$ then $\theta_k=w$ $\forall k>0$ including no association when $w=0$.  

We assign a prior to $\theta_k$, $k=1,\dots,M$, that is a finite mixture of a mass point at zero denoted by the Dirac measure $\delta_0$ and a distribution $P$ with positive support. This approach selects some regression coefficients to be 0, effectively removing those basis functions from the model. In the non-zero probability event that all regression coefficients are zero there is no association between $x$ and $y$. We then let the positive distribution be a Dirichlet process (DP) with base measure $P_0\equiv \text{TN}_{[0,\infty]} (\mu,\phi^2)$, where $\text{TN}_{[0,\infty]}(\mu,\phi^2)$ implies a truncated normal with support $[0,\infty]$, mean $\mu$, and variance $\phi^2$. By using a base measure with support over $\mathbb{R}^+$ we ensure that the non-zero regression coefficients are positive. This imposes monotonicity of $f$. Further, the clustering property of the DP allows for all regression coefficients to be equal, in the same cluster, allowing for positive probability that $f$ is linear. The selection and clustering of the regression coefficients does not, however, impact smoothness.  The estimated function is guaranteed to be smooth and differentiable.

The full model is
\begin{eqnarray}
\label{eq:fullmodel}
Y_i| \boldsymbol\theta,\sigma^2 &\sim& \text{N}\left[ \Psi(x_i,M) \mathbf{A}^{-1} \boldsymbol\theta , \sigma^2 \right] \\
\theta_j| P,\pi &\sim& \pi \delta_0 + (1-\pi)P\nonumber \\ 
P &\sim& \text{DP}(\alpha P_0)\nonumber\\
P_0 &\equiv& \text{TN}_{[0,\infty]} (\mu,\phi^2). \nonumber
\end{eqnarray}
The above model is equivalent to a DP with base measure that is a finite mixture 
\begin{eqnarray}
\theta_j|G &\sim&G \\ 
G &\sim& \text{DP}(\alpha G_0)\nonumber\\
G_0|\pi &\equiv&\pi \delta_0 + (1-\pi)\text{TN}_{[0,\infty]} (\mu,\phi^2). \nonumber
\end{eqnarray}
 Several papers have used similar DP constructions that combine a DP with a finite mixture of a mass point and a non-truncated normal distribution \citep{Herring2010,Canale2017,Cassese2019} or a gamma distribution \citep{Liu2015a}. 

We complete the specification by assigning the prior $\sigma^{-2}\sim\text{Gamma}(a,b)$, a normal mean zero variance $\phi_0^2$ prior to the intercept $\theta_0$, and $\pi\sim\text{Beta}(a_\pi,b_\pi)$.

\subsection{Posterior computation}

The model in \eqref{eq:fullmodel} can be efficiently sampled with a Gibbs sampler. This is accomplished by first integrating out $\pi$ from the model. The Gaussian likelihood and truncated normal base measure allows for $P$ to be marginalized out of the model as well. The posterior can be simulated using a Polya Urn scheme \citep{Blackwell1973,West1994,Bush1996}. 

Let $\Lambda_i=\Psi(x_i,M)\mathbf{A}^{-1}$ be the transformed BP basis expansion for observation $i$ and $\boldsymbol\Lambda$ be the $n\times (M+1)$ design matrix with row $i$ equal to $\Lambda_i$. Let $\Lambda_{i[k]}$ denote the vector $\Lambda_i$ with the $k^{th}$ element omitted, $\boldsymbol\theta_{[k]}$ the vector $\boldsymbol\theta$ with the $k^{th}$ element omitted, and $\Lambda_{ik}$ denote only the $k^{th}$ element of $\Lambda_i$. Similarly, let $\boldsymbol\Lambda_{[k]}$ be the matrix $\boldsymbol\Lambda$ with the $k^{th}$ column omitted and $\boldsymbol\Lambda_{k}$ be only the $k^{th}$ column of $\boldsymbol\Lambda$. Finally, we denote by $S_k$ the categorical indicator where $S_k=c$ if $\theta_k=\eta_c$ and $n_c$  the number of coefficients in cluster $c$ where $n_0$ is the number in the null cluster with $\theta_k=0$. 

The full conditional for $S_k$, $k=1,\dots,S_M$, is categorical. The conditional probability that the $k^{th}$ regression coefficient is equal to 0 is
\begin{equation}
\Pr\left(S_k=0|-\right) = d \frac{n_0^*+a_\pi}{M-1+a_\pi + b_\pi} \prod_{i=1}^n\mathcal{N}\left(y_i; \Lambda_{i[k]} \boldsymbol\theta_{[k]},\sigma^2\right),
\end{equation}
where $n_0^*$ is the number of regression coefficients in cluster $0$, the null cluster, excluding $\theta_k$, $d$ is a normalizing constant, and $\mathcal{N}(x;\mu,\sigma^2)$ denotes a normal density function. In contrast to standard DP models, the zero cluster is allowed to be empty in this model. The conditional probability that $\theta_k$ is allocated to an existing non-zero cluster $c$ is 
\begin{eqnarray}
\Pr\left(S_k=c|-\right) &=& d \frac{\left(M-n_0^*-1+b_\pi\right)n_c^*}{\left(M-1+a_\pi + b_\pi\right)\left(n-n_0^*-1+\alpha\right)} \\
&&\quad\times\prod_{i=1}^n\mathcal{N}\left(y_i; \Lambda_{i[k]} \boldsymbol\theta_{[k]}+\Lambda_{ik} \eta_c,\sigma^2\right).\nonumber
\end{eqnarray}
Finally, the conditional probability that $\theta_k$ is allocated to a new cluster $c'$ is
\begin{eqnarray}
\Pr\left(S_k=c'|-\right) &=& d \frac{\left(M-n_0^*-1+b_\pi\right)\alpha}{\left(M-1+a_\pi + b_\pi\right)\left(M-n_0^*-1+\alpha\right)} \\
&&\quad\times\left[\prod_{i=1}^n\mathcal{N}\left\{y_i;\sum_{l=0,l\ne k}^M\phi_l(x_i,M)\beta_l,\sigma^2\right\}\right] \exp\left(\frac{\widetilde{m}^2}{2\widetilde{v}}-\frac{\mu^2}{2\phi^2}\right)\nonumber\\
&&\quad\times
\frac{(2\pi)^{-1/2} \phi^{-1} \widetilde{v}^{1/2}}{\int_0^\infty f(z;\mu,\phi^2)dz } 
\int_0^\infty f(\theta;\widetilde{m},\widetilde{v})d\theta.\nonumber
\end{eqnarray}
where $\widetilde{v}=1/[\phi^{-2} + \sigma^{-2}\sum_{i=1}^n\psi_k(x_i,M)^2]$ and $\widetilde{m}=\widetilde{v}[\phi^{-2}\mu+ \sigma^{-2}\sum_{i=1}^n\psi_k(x_i,M)\{y_i-\sum_{l=0,l\ne k}^M\psi_l(x_i,M)\beta_l\}]$.

In the situation where $S_k$ is assigned to new cluster $c'$ a value for $\theta_k=\eta_{c'}$ can be sampled from its univariate truncated normal full conditional. The full conditional for a single regression coefficient $\eta_{c'}$ where $n_{c'}=1$ (no other coefficient takes that value) is truncated above 0 and has mean $\widetilde{m}$ and variance $\widetilde{v}$ as specified above. We use the hybrid univariate truncated normal sampler of \cite{Li2015} to sample from this full conditional.

The $M+1$-vector $\boldsymbol\theta$ contains three types of elements: the unconstrained intercept, parameters that are selected to be 0, and parameters that are non-zero and are constrained to be greater than 0. The non-zero values take on $K+1$ unique values $\boldsymbol\eta=\{\theta_0,\eta_1,\dots,\eta_K\}$ where $\theta_0$ is the unconstrained intercept. Using this notation the linear predictor $\boldsymbol\theta=\mathbf{B}\boldsymbol\eta$ where $\mathbf{B}$ is a transformation matrix that maps $\boldsymbol\eta$ to $\boldsymbol\theta$ according to $S_1,\dots,S_M$. The vector $\boldsymbol\eta$ has a truncated multivariate normal full conditional with mean $m=\sigma^{-2}v\left(\mathbf{B}^T\boldsymbol\Lambda^T\mathbf{y}+\phi^{-2}\mu\right)$ and variance $v=\left(\sigma^{-2}\mathbf{B}^T\boldsymbol\Lambda^T\boldsymbol\Lambda\mathbf{B} + D\right)$ where $D$ is is a diagonal matrix with $\phi_0^{-2}$ in the first diagonal location for the intercept and $\phi^{-2}$ in all other diagonal locations for the constrained coefficients. These are the same as the typical mean and variance for a normal-normal model full conditional. The first element $\theta_0$ is not truncated and the remaining elements are truncated below at 0. We simulate from the full conditional of $\boldsymbol\eta$ as a multivariate block using the hybrid multivariate sampler approach of \cite{Li2015}.

The Gibbs sampler is completed with standard updates of $\alpha$ using a mixture of gammas \citep{Escobar1995} and $\sigma^{-2}$ using the standard gamma full conditional.

\subsection{Details on tuning}

Care must be given when specifying the prior, particularly for the choice of values for the mean and standard deviation of the base measure $\mu$ and $\phi$. This is challenging because the plausible values for the regression coefficients depends on the number of non-zero regression coefficients in the model and how many basis functions each coefficient is applied to (cluster size). We do not know either of these quantities a priori. We have taken the approach of scaling the outcome $\mathbf{y}$ to have mean zero and variance one and then setting $\mu=0.5$ and $\phi=0.25$. This puts reasonable mass on values between zero and one which represents plausible values for a variety of basis configurations. We have found that this choice performs well across a variety of simulated and real datasets. We use this setting in all simulation and data analysis results presented in this paper. However, results can be sensitive to this choice. Supplemental Section 1.2 includes an additional simulation study that compares sensitivity to different values of $\mu$ and $\phi$. We show that as $\phi$ increases the posterior probability of no association decreases  and the number of clusters (unique non-zero regression coefficients) increases. However, the model fit as measured by RMSE on $f$ and the derivative of $f$ is less sensitive to this choice.

The user must also specify the order of the BP ($M$). This should be selected, in theory, based on sample size and the differentiability of the function being estimated \citep{Mclain2009}. In practice, methods such as reversible jump MCMC or Kullback-Leibler distance have been used to attempt to estimate the dimension of basis expansions to be used in nonparametric regression \citep[e.g.][]{Dias2002,Dias2007,Meyer2011} while penalization can be used to regularize a rich basis to avoid over fitting \citep{Crainiceanu2005}. It has additionally been noted that shape constraints, including monotonicity, reduce sensitivity to the dimension of the basis expansion \citep{Meyer2008}. We follow the approach of \cite{Curtis2011} and use a rich basis and let the prior select or cluster redundant predictors. In this paper, we use $M=50$ in all results shown in the main text but show in the supplement that a smaller value of $M$ results in lower RMSE when the true function is closer to linear and a higher value of $M$ is preferred when the true function is more wiggily. If the practitioner has prior knowledge of the shape of the underlying function, beyond monotonicity, this could be incorporated into the selection of $M$ a priori.

\subsection{Inference on the derivative and aerosol concentration}\label{sub:derivative}

The proposed approach allows for coherent estimation and inference on not only the function $f$ but on the derivatives of $f$. This includes full quantification of the uncertainty in the derivatives and guaranteed smoothness in the derivatives. This is particularly critical in our application where the first derivative of $f$ is proportional to the time-resolved aerosol concentration. For a BP of order $M$ the first derivative is a BP of order $M-1$.  Specifically the first derivative is
\begin{equation}
    f'(x) = M\sum_{k=0}^{M-1}\psi_{k}(x,M-1)\theta_{k+1} = M\Psi(x,M-1)\boldsymbol\theta_{[0]}.
    \label{eq:deriv1}
\end{equation}
For the derivative the regression coefficient $\theta_0=\beta_0$, which corresponds to the intercept, is not included. Hence, the derivative can be identified in closed form from the posterior sample of $\boldsymbol\theta$. Inference on the derivative can be made directly by using the posterior sample of $\boldsymbol\theta$.

From a theoretical perspective the total aerosol mass accumulated on the filter should be the flow rate through the filter times the concentration integrated over time. Here, flow rate is constant and therefore $\int_0^1 f'(x) dx \propto (\text{filter mass})/(\text{flow rate})$.  In our model $\int_0^1 f'(x) dx = \beta_M-\beta_0=\sum_{k=1}^M\theta_k$. We therefore scale the derivative to the total filter mass by replacing $\boldsymbol\theta$ in \eqref{eq:deriv1} with $\tilde{\boldsymbol\theta}=\boldsymbol\theta\times(\text{filter mass})/[(\text{flow rate})\times\sum_{k=1}^M\theta_k]$. We then estimate aerosol concentration as
\begin{equation}
    \tilde{f}'(x) = M\sum_{k=0}^{M-1}\psi_{k}(x,M-1)\tilde\theta_{k+1} = M\Psi(x,M-1)\tilde{\boldsymbol\theta}_{[0]}.
    \label{eq:deriv2}
\end{equation}
In practice, we scale each draw of $\boldsymbol\theta$ from the posterior and then construct a  posterior sample of $\tilde{f}'$.

\subsection{Alternative spline approach}
The proposed prior can be applied to other basis expansions and achieve some, but not all, of the same properties. Using the same prior structure with a transformed $B$-spline or $I$-splines without the transformation matrix $\mathbf{A}$ can still achieve monotonicity \citep{DeBoor1978,Ramsay1988}.  The proposed prior will also allow estimation of no association when all regression coefficients are clustered at zero. However, the clustering will not result in shrinkage toward a linear response. 

Using a spline basis with compact support may result in more flexibility than the BP approach presented here. This could be particularly appealing when the function being estimated has sharp change points. In addition, the derivative of many common splines including $B$-splines and $I$-splines can be represented as a spline itself and inference on the derivative can be made using a similar approach. However, splines lose flexibility and smoothness in the derivative. For example, the standard cubic spline has a quadratic derivative while a quadratic spline has a piecewise linear derivative. This may not be sufficiently flexible in many cases, as seen in the data analysis in Section~\ref{s:da}. In contrast, the BP uses a higher order polynomial and therefore has a higher order derivative which imposes smoothness not only in the function being estimated but in all derivatives of that function.

\section{Simulation}

We compare the proposed approach, BNMR, to alternative methods for monotone regression in a simulation study. We generated 500 datasets from four designs each taking the form $y_i=f_s(x_i)+\epsilon_j$ for $i=1,\dots,n$ with $\epsilon_i\sim\text{N}(0,0.25^2)$. We generate $x\sim\text{Unif}(0,1)$ and consider four shapes for the function $f_s(\cdot)$:
\begin{enumerate}
\item Flat: $f_1(x)=0$.
\item Linear: $f_2(x)=x$.
\item Wavy: $f_3(x)=\sin(3\pi x)/(3\pi)+x$.
\item Flat-nonlinear: $f_4(x)=0$ for $x<0.5$ and $f_5(x)=[2(x-0.5)]^2$ for $0.5\le x$.
\end{enumerate}
The flat, linear, and wavy functions mirror those from \cite{Curtis2011}. We simulated data sets of size $n=100$ and $1000$.

We compared BNMR to alternative monotone regression methods that have available {\tt R} software that includes variance estimates. The comparison methods are: constrained generalized additive models \citep[CGAM,][]{Meyer2013,Meyer2018}, Bayesian constrained generalized additive models \citep[BCGAM,][]{Meyer2011,Oliva-Aviles2018}, and Bayesian isotonic regression \citep[BISOREG,][]{Curtis2011,Curtis2018}. In addition we compare with the unconstrained methods ordinary least squares (OLS), local polynomial regression (LOESS), and an unconstrained Bernstein polynomial model (UBP). For BNMR and BISOREG we set $M=50$ and consider other values in the Supplemental Material. For UBP we select $M$ using deviance information criterion  \citep[DIC,][]{Spiegelhalter2002}.

We evaluate the model performance by the root mean squared error (RMSE) on the function $f(\cdot)$ and the pointwise 95\% interval coverage both evaluated at 100 evenly spaced points spanning the range of $x$. For the Bayesian methods (BNMR, BISOREG, BCGAM, and UBP) we consider the posterior probability that $f$ is linear and that $f$ is flat (no association). For the CGAM and OLS we report the mean $p$-value for testing the null hypothesis that there is no association. 

Table~\ref{tab:sim1} shows results from the simulation study. At $n=100$, BNMR had the lowest RMSE on $f$ among all the monotone regression methods on all four scenarios (within one standard error with BCGAM and BISOREG on the flat scenario). The only method to have lower RMSE was OLS on the linear scenario. BNMR, BISOREG, CGAM, and UBP all had pointwise 95\% interval coverage between 0.95 and 0.98 on all scenarios. Each of the other methods had interval coverage of 0.9 or less on at least one scenario. At $n=1000$, BNMR had the lowest RMSE for the flat and flat-nonlinear scenarios (within one standard error with BCGAM and BISOREG on the flat scenario). CGAM and BCGAM had the lowest RMSE of the constrained methods on the wavy scenario with BNMR and BISOREG slightly higher. OLS and UBP had the lowest RMSE on the linear scenario. UBP selected $M=3$, a cubic regression function, in almost all datasets for the linear scenario.

\begin{table}
\centering
\caption{ Simulation results comparing estimation of $f$ with each method. The table shows RMSE and 95\% interval coverage both evaluated pointwise on a grid of 100 evenly spaced points. The columns labeled Pr( flat ) are the posterior probability of a flat response or for OLS and CGAM the mean $p$-value for rejecting the null of association. Additional simulation results including standard errors for the RMSE and interval widths are included in the supplemental material.}
\label{tab:sim1}
\begin{tabular}{lcccccc}\hline
& \multicolumn{3}{c}{$n=100$} & \multicolumn{3}{c}{$n=1000$}\\ \cmidrule(lr){2-4} \cmidrule(lr){5-7}
Model	&	 RMSE &	Coverage &	Pr( flat ) &	 RMSE &	Coverage &	Pr( flat )	\\ \hline
\multicolumn{7}{l}{\it Scenario 1: Flat}\\				
BCGAM & 2.22 & 0.95 & 0.00 & 0.65 & 0.96 & 0.00  \\ 
BISOREG & 2.19 & 0.97 & 0.86 & 0.61 & 0.97 & 0.95  \\ 
BNMR & 2.08 & 0.96 & 0.94 & 0.60 & 0.96 & 0.99  \\ 
CGAM & 3.57 & 0.98 & 0.49 & 1.17 & 0.99 & 0.50  \\ 
LOESS & 5.15 & 0.93 &   NA & 1.61 & 0.95 &   NA  \\ 
OLS & 3.11 & 0.95 & 0.50 & 0.93 & 0.96 & 0.53  \\ 
UBP & 5.23 & 0.93 & 0.00 & 1.60 & 0.95 & 0.00  \\  \hline
\multicolumn{7}{l}{\it Scenario 2: Linear}\\	
BCGAM & 6.42 & 0.84 & 0.00 & 2.58 & 0.84 & 0.00  \\ 
BISOREG & 5.99 & 0.96 & 0.00 & 2.42 & 0.95 & 0.00  \\ 
BNMR & 4.72 & 0.98 & 0.00 & 2.14 & 0.96 & 0.00  \\ 
CGAM & 5.79 & 0.96 & 0.00 & 2.30 & 0.95 & 0.00  \\ 
LOESS & 5.15 & 0.93 &   NA & 1.61 & 0.95 &   NA  \\ 
OLS & 3.11 & 0.95 & 0.00 & 0.93 & 0.96 & 0.00  \\ 
UBP & 5.28 & 0.93 & 0.00 & 1.60 & 0.95 & 0.00  \\  \hline	
\multicolumn{7}{l}{\it Scenario 3: Wavy}\\
BCGAM & 6.57 & 0.84 & 0.00 & 2.17 & 0.88 & 0.00  \\ 
BISOREG & 5.98 & 0.95 & 0.00 & 2.25 & 0.95 & 0.00  \\ 
BNMR & 5.30 & 0.96 & 0.00 & 2.25 & 0.94 & 0.00  \\ 
CGAM & 5.67 & 0.96 & 0.00 & 2.15 & 0.96 & 0.00  \\ 
LOESS & 6.44 & 0.89 &   NA & 2.14 & 0.93 &   NA  \\ 
OLS & 8.02 & 0.56 & 0.00 & 7.22 & 0.19 & 0.00  \\ 
UBP & 6.38 & 0.90 & 0.00 & 2.30 & 0.90 & 0.00  \\  \hline	
\multicolumn{7}{l}{\it Scenario 4: Flat-nonlinear}\\
BCGAM &  5.33 & 0.90 & 0.00 &  1.93 & 0.89 & 0.00  \\ 
BISOREG &  5.60 & 0.95 & 0.00 &  2.12 & 0.96 & 0.00  \\ 
BNMR &  4.95 & 0.96 & 0.00 &  1.82 & 0.96 & 0.00  \\ 
CGAM &  5.29 & 0.96 & 0.00 &  1.91 & 0.97 & 0.00  \\ 
LOESS &  5.70 & 0.91 &   NA &  1.88 & 0.93 &   NA  \\ 
OLS & 16.11 & 0.32 & 0.00 & 16.24 & 0.09 & 0.00  \\ 
UBP &  5.42 & 0.93 & 0.00 &  1.91 & 0.91 & 0.00  \\ \hline	
\end{tabular}
\end{table}

Both BNMR and BISREG had high, greater than 0.86, posterior probabilities of a flat response (no association) in the flat scenario. BCGAM does not include a flat response in the parameter space and therefore has a posterior, and prior, probability of 0. The average $p$-values for the test of no association for CGAM and OLS were between 0.49 and 0.53. 

\begin{figure}
    \centering
    \includegraphics{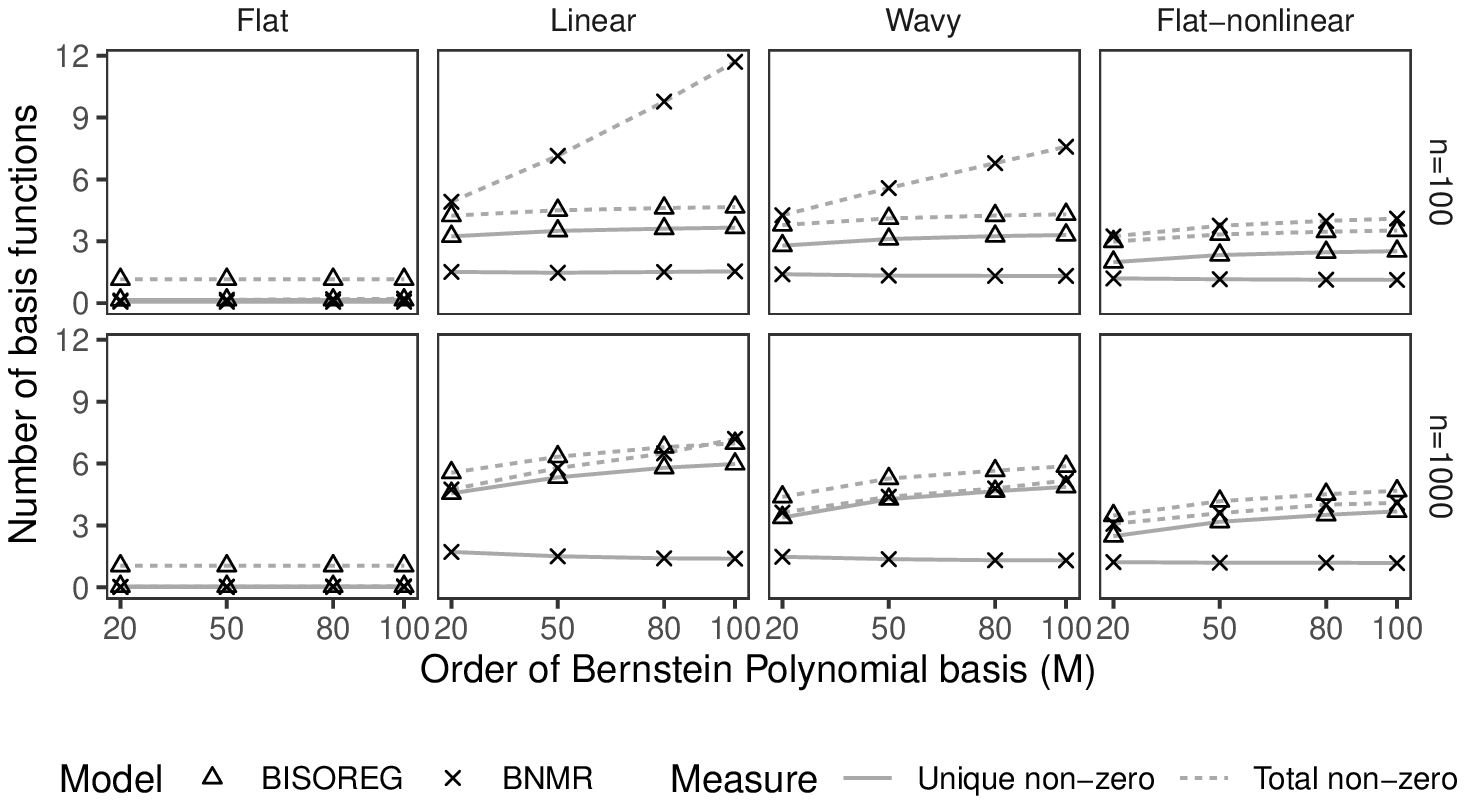}
    \caption{Simulation results for the number of non-zero regression coefficients (dashed line) and the number of unique values of the non-zero regression coefficient (solid line) for BISOREG (triangle) and BNMR ($\times$). The number of unique non-zero values is always equal to the total number of non-zero values in BISOREG.}
    \label{fig:nparams}
\end{figure}

BNMR is the only method that allows the estimated function $f$ to be linear with slope greater than 0. However, in the linear scenario this did not occur. The mean posterior probability of a linear function was 0.00. The response is linear when all regression coefficients are non-zero and take the same value.  Figure~\ref{fig:nparams} shows the number of non-zero regression coefficient in the model and the number of unique values those regression coefficients take.  Both BNMR and BISOREG include only a small number of non-zero regression coefficients, effectively selecting out of the model the majority of the basis functions. Because not all basis functions are included the estimated regression function is never truly linear. Despite not being exactly linear, BNMR has lower RMSE than any of the other nonparametric methods on the linear scenario.

A key difference between BNMR and BISOREG is that all non-zero regression coefficients in BISOREG take unique values while with BNMR the non-zero regression coefficients are clustered together and take fewer unique values. On average, there were more non-zero regression coefficients included into the model with BNMR but fewer, less than two, unique regression coefficients. This is true for both the linear and nonlinear scenarios and for BP expansions of order ranging from 20 to 100 (shown in supplemental material). As a result, the proposed approach requires estimating only a small number of unique regression coefficients regardless of the size of the basis expansion or the wiggliness of the regression function.

In our application we are interested in the derivative of the monotone function.  The BP basis used by BISOREG, BNMR and UBP allows straight forward inference on the derivatives of $f$. The other methods do not naturally allow for this inference. We calculate a pointwise approximation of the derivative for the other methods by calculating change in $\widehat{f}$ divided by change in $x$ for each pair of neighboring observations on an equally spaced grid. We do not evaluate coverage for these methods. Table~\ref{tab:sim1deriv} shows the RMSE and 95\% interval coverage (for BISOREG, BNMR and UBP only) for the derivative of $f$. 
BNMR had lowest RMSE for all scenarios at the smaller sample size and the flat scenario at the larger sample size. BNMR, BISOREG, and UBP all suffered from poor interval coverage in several of the scenarios. The coverage is pointwise and in the flat-nonlinear scenario, which has an sharp ``elbow'' change-point, both methods fail to cover in the elbow, highlighting a limitation of the smooth BP basis.

\begin{table}
\centering
\caption{ Simulation results comparing estimation of the derivative $f'$ with each method. The table shows RMSE and 95\% interval coverage both evaluated pointwise on a grid of 100 evenly spaced points. Intervals for the derivative with BCGAM, CGAM and LOESS are not available. Additional simulation results including standard errors for the RMSE and interval widths are included in the supplemental material.}
\label{tab:sim1deriv}
\begin{tabular}{lcccc}\hline
& \multicolumn{2}{c}{$n=100$} & \multicolumn{2}{c}{$n=1000$}\\ \cmidrule(lr){2-3} \cmidrule(lr){4-5}
Model	&	 RMSE &	Coverage  &	 RMSE &	Coverage 	\\ \hline
\multicolumn{5}{l}{\it Scenario 1: Flat}\\				
BCGAM &  3.24 &   NA &  0.85 &   NA  \\ 
BISOREG &  4.90 & 0.00 &  0.61 & 0.00  \\ 
BNMR &  1.12 & 1.00 &  0.19 & 1.00  \\ 
CGAM & 22.80 &   NA &  9.92 &   NA  \\ 
LOESS & 53.81 &   NA & 18.49 &   NA  \\ 
UBP & 55.26 & 0.92 & 16.61 & 0.93  \\  \hline
\multicolumn{5}{l}{\it Scenario 2: Linear}\\	
BCGAM & 61.68 &   NA & 39.98 &   NA  \\ 
BISOREG & 65.54 & 0.96 & 44.91 & 0.94  \\ 
BNMR & 39.57 & 1.00 & 38.21 & 0.95  \\ 
CGAM & 68.85 &   NA & 40.79 &   NA  \\ 
LOESS & 53.81 &   NA & 18.49 &   NA  \\ 
UBP & 56.90 & 0.91 & 16.60 & 0.93  \\  \hline	
\multicolumn{5}{l}{\it Scenario 3: Wavy}\\
BCGAM & 64.03 &   NA & 32.05 &   NA  \\ 
BISOREG & 70.78 & 0.95 & 46.53 & 0.91  \\ 
BNMR & 55.00 & 0.97 & 45.13 & 0.85  \\ 
CGAM & 69.21 &   NA & 37.04 &   NA  \\ 
LOESS & 87.54 &   NA & 35.18 &   NA  \\ 
UBP & 97.38 & 0.78 & 52.52 & 0.68  \\ \hline	
\multicolumn{5}{l}{\it Scenario 4: Flat-nonlinear}\\
BCGAM & 62.08 &   NA & 28.82 &   NA  \\ 
BISOREG & 92.18 & 0.46 & 65.44 & 0.45  \\ 
BNMR & 61.45 & 0.68 & 46.91 & 0.61  \\ 
CGAM & 66.95 &   NA & 34.57 &   NA  \\ 
LOESS & 65.11 &   NA & 28.99 &   NA  \\ 
UBP & 62.59 & 0.88 & 28.39 & 0.66  \\  \hline				
\end{tabular}
\end{table}

The supplemental material includes additional simulation results, including standard errors for the estimates in Tables~\ref{tab:sim1} and \ref{tab:sim1deriv}, interval widths, and results on sensitivity to the choice of prior $\mu$ and $\phi^2$ as well as the order of the BP $M$. In addition we show results for computation time as a function of sample size and order of the BP.

\section{Analysis of Real-Time PM$_{2.5}$ Concentration Inferred from Pressure Drop}\label{s:da}

\subsection{Overview of the data analysis}
We use data from 12 samples collected using three MARS devices during four laboratory experiments. These experiments are described in detail by \citet{Tryner2019a}. During each experiment, one of four different types of aerosol---urban PM (NIST SRM 1648A Urban PM), ammonium sulfate ((NH$_4$)$_2$SO$_{4}$), Arizona road dust, or match smoke---is nebulized into a controlled chamber containing all three MARS. Each MARS samples PM$_{2.5}$ onto a new polytetrafluoroethylene (PTFE) filter at a flow rate of 1 L min$^{-1}$ for between 7.5 and 13 hours while pressure drop across the filter is recorded every 30 seconds. Each filter is weighed before and after the experiment to measure the total mass of PM$_{2.5}$ accumulated. A TEOM measures the PM$_{2.5}$ concentration in the chamber every minute as a previously-validated point of comparison.

We use BNMR to estimate time-resolved PM$_{2.5}$ concentration using the MARS pressure drop data from the 12 samples. Prior to analysis we removed the first 30 minutes of pressure drop as: 1) there was no PM$_{2.5}$ in the chamber at that time and 2) the new filter was stretching during that time and a decreasing trend is observed due to the stretching process. We also removed the final five minutes when there was 1) no PM$_{2.5}$ in the chamber and 2) the sampler was shutting down resulting in spurious noise in the pressure drop function. Then, we fit BNMR to the time-series of measured pressure drop for each sample. From the fitted model we then estimate the scaled first derivative of pressure drop at each time-point for which the TEOM recorded PM$_{2.5}$ as described in Section~\ref{sub:derivative}. For comparison, we perform the same procedure with BISOREG and UBP.  We also estimate pressure drop and the scaled pointwise approximation of the derivative with LOESS, CGAM, and BCGAM. We omit OLS because of obvious nonlinearities in the pressure drop data.

For each method we visualize and compare the performance with respect to estimating the pressure drop function and inferring real-time PM$_{2.5}$ from the scaled derivative of pressure drop. We show results from one of the 12 samples in the main text. The supplemental material includes estimated pressure drop and estimated real-time PM$_{2.5}$ concentration for all 12 samples.

\subsection{Estimation of the pressure drop function}

\begin{figure}
    \centering
    \subfloat[UBP]{
    \includegraphics[]{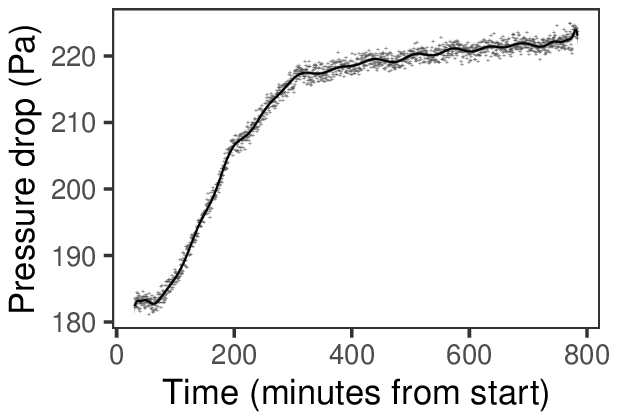}
    \label{subfig:ubp}
    }
    \subfloat[BCGAM]{
    \includegraphics[]{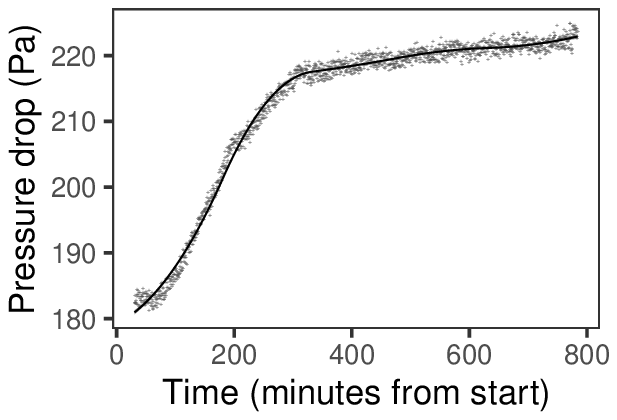}
    \label{subfig:bcgam}
    }
    
    \subfloat[BNMR]{
    \includegraphics[]{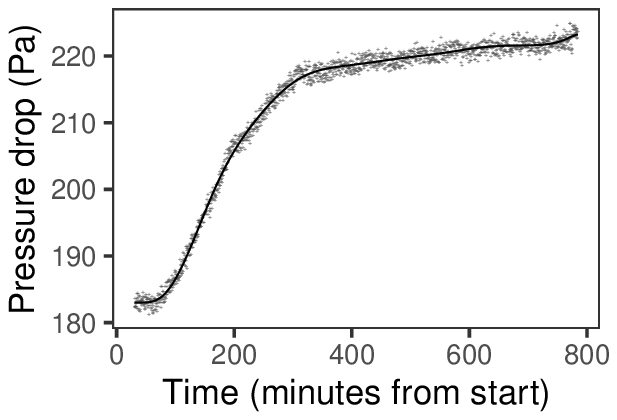}
    \label{subfig:bnmr}
    }
    \subfloat[BISOREG]{
    \includegraphics[]{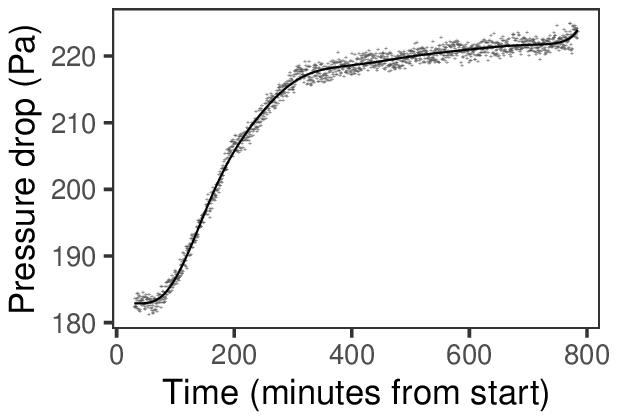}
    \label{subfig:bisoreg}
    }
    
    \subfloat[CGAM]{
    \includegraphics[]{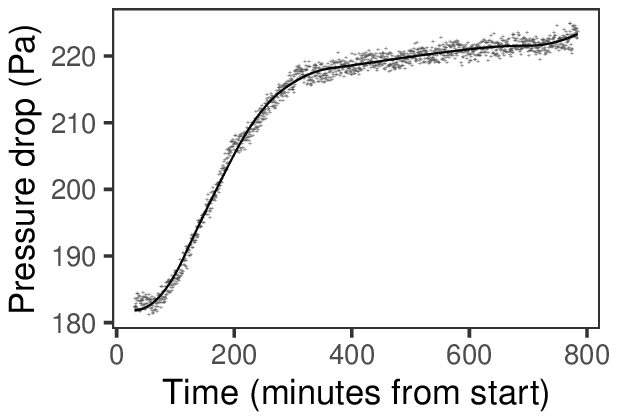}
    \label{subfig:cgam}
    }
    \subfloat[LOESS]{
    \includegraphics[]{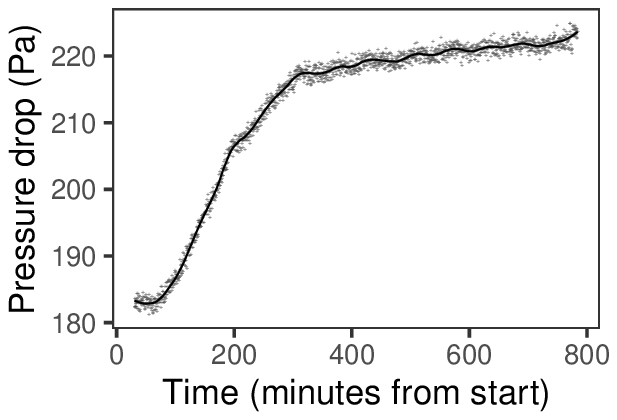}
    \label{subfig:loess}
    }
    \caption{Estimated pressure drop from the MARS data for one run. Each panel shows the estimates and 95\% intervals for each method separately. Results from all 12 runs are shown in the supplemental material.}
    \label{fig:analysis1}
\end{figure}

Figure~\ref{fig:analysis1} shows the data and estimates from all six methods for a single sample. Each panel show estimates from a single method along with 0.95 confidence or credible intervals. The fits are near identical visually over most of the range. However, there are differences in the lower tail.  BNMR and BISOREG tend to level-off between 0 and 100 minutes. In contrast CGAM and especially BCGAM tend to over-smooth over the same time period. 

Comparing UBP to BNMR and BISOREG, which all use a BP basis, highlights an important difference between the constrained methods and the unconstrained method. Specifically, UBP experiences instability in the tails, while BNMR and BISOREG which impose monotonicity and further regulate with a selection prior (BISOREG) or a selection and clustering prior (BNMR), are more stable in the tails. 

To formally compare model fit for the pressure drop function we performed five-fold cross-validation for each sample. Table~\ref{tab:cv} shows cross-validation results for all five methods across all 12 samples. LOESS had the lowest cross-validation RMSE at 0.81 followed by UBP at 1.21. Hence, the unconstrained methods provided a better fit then any of the monotone methods. The best performing monotone methods were BISOREG at 1.27 and BNMR 1.29. CGAM and BCGAM had higher RMSEs of 1.47 and 1.96, respectively.

\begin{table}
\centering
\caption{Summary of the model fit for each method in the data analysis. The table shows the cross-validation RMSE from the five-fold cross validation. For BNMR and BISOREG the table additionally shows the comparison of the scaled derivative to the time-resolved measurement of PM$_{2.5}$ from a TEOM. The results show the $R^2$, intercept, and slope from the regression of the TEOM PM$_{2.5}$ on the MARS estimated PM$_{2.5}$ obtained from the estimated first derivative of pressure drop. }
\label{tab:cv}
\begin{tabular}{lcccc}\hline
& \multicolumn{1}{c}{Cross-Validation} & \multicolumn{3}{c}{Regression}\\ \cmidrule(lr){2-2} \cmidrule(lr){3-5}
Model	&	 RMSE &	$R^2$  &	 Intercept &	Slope 	\\ \hline
BCGAM & 1.96 & 0.57 & 37.27 & 0.97  \\ 
BISOREG & 1.27 & 0.75 & 46.89 & 0.91  \\ 
BNMR & 1.29 & 0.75 & 44.59 & 0.92  \\ 
CGAM & 1.47 & 0.72 & -2.03 & 1.09  \\ 
LOESS & 0.81 & 0.81 & 51.60 & 0.88  \\ 
UBP & 1.21 & 0.63 & 84.77 & 0.69  \\ 
\hline
\end{tabular}
\end{table}

LOESS outperforms the other methods in terms of cross-validation RMSE on the pressure drop function for two reasons. First, LOESS does not impose monotonicity and several of the pressure drop measurements show minor deviations from the largely monotone trend. The small waves result from small fluctuations in the air temperature measured by the device, which lead to small fluctuations in air density and thus small fluctuations in the mass flow rate through the filter. The second reason that LOESS has lower cross-validation RMSE is that three of samples show sharp change-points in the pressure drop functions (similar to the ``elbow'' in simulation scenario 4) and LOESS is the only method that did not over-smooth these points (see supplemental Figure 9). UBP can also estimate the non-monotone trend but struggles with the ``elbow.'' However, the non-monotonicity in LOESS and UBP results in negative estimates of aerosol concentration, which are not physically possible.  

The monotone methods smooth over the non-monotone areas of the data. This results in valid estimates of PM$_{2.5}$ because the derivative is always non-negative. It is also consistent with the theoretical framework for measuring time-resolved PM$_{2.5}$ from pressure drop using MARS as the pressure drop function should be monotone. However, this comes at a cost because the oscillation appears as autocorrelation in the residuals. This is not accounted for in the model as we assume independent and identically distributed residuals and could result in some bias in the intervals but results in a rational estimate of time-resolved PM$_{2.5}$.

\subsection{Inference on time-resolved PM$_{2.5}$ with the scaled derivative}

\begin{figure}
    \centering
    \subfloat[BNMR]{
    \includegraphics[]{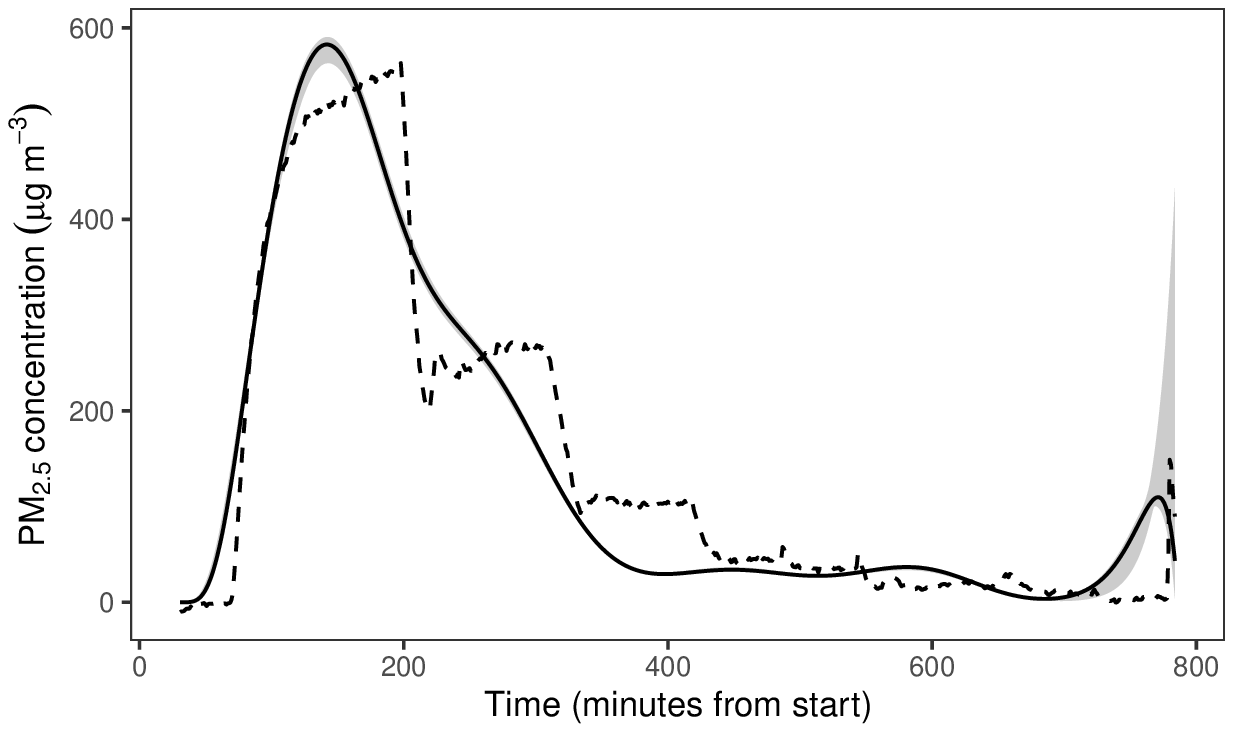}
    \label{subfig:bnmr2}
    }
    
    \subfloat[BISOREG]{
    \includegraphics[]{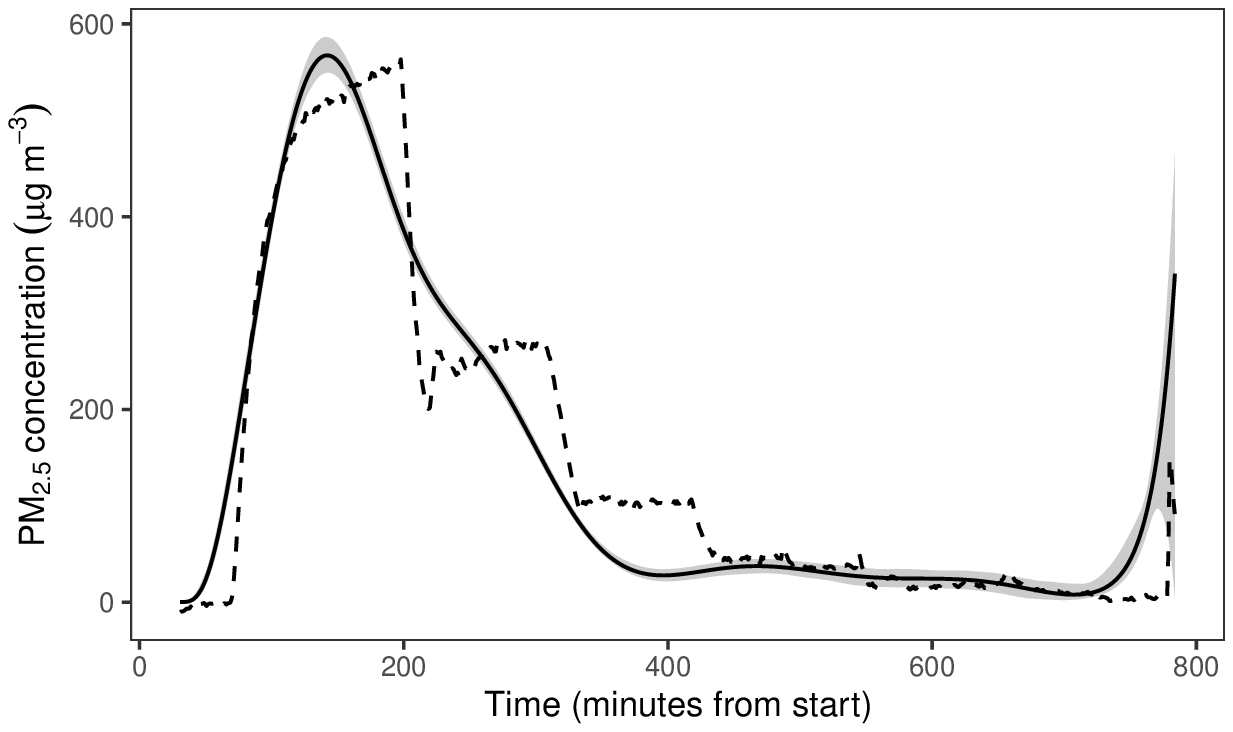}
    \label{subfig:bisoreg2}
    }

    \caption{Estimated PM$_{2.5}$ concentration from the MARS data.  Panel~\ref{subfig:bnmr2} shows the posterior mean and 95\% interval from BNMR and Panel~\ref{subfig:bisoreg2} shows the posterior mean and 95\% interval from BISOREG. The dashed line in each panel is the PM$_{2.5}$ concentration measured with the TEOM. Results from all 12 runs are shown in the supplemental material. Results with other methods are also shown in the supplemental material.}
    \label{fig:analysis2}
\end{figure}

Our primary interest is estimating PM$_{2.5}$ concentration using the scaled first derivative of the estimated pressure drop function. We scale each derivative by the total mass collected on the filter as described in Section~\ref{sub:derivative}. Figure~\ref{fig:analysis2} shows the estimates for the scaled first derivative with both BNMR and BISOREG. For comparison, the PM$_{2.5}$ concentration measured with the TEOM is included. Both BNMR and BISOREG estimate the larger pattern in PM$_{2.5}$ concentration but do not fully capture localized features.

To more formally compare the estimates of PM$_{2.5}$ concentration we regressed the one minute TEOM measurements on the estimated concentrations at those same time points obtained using each method (Table 3).  The mean $R^2$ across all 12 samples was 0.75 with BNMR and BISOREG. Hence, these two approaches provide similar estimates of real-time concentration. 

Estimates of PM$_{2.5}$ from the other methods (BCGAM, CGAM, LOESS, and UBP) are presented in the supplement. All of these methods are being used beyond their original intention and suffer from shortcomings when estimating the derivative of a function. BCGAM and CGAM use a quadratic spline and result in piecewise linear derivatives which are not suitable to estimate the time-resolved PM$_{2.5}$. LOESS and UBP are non-monotone and result in negative estimates of PM$_{2.5}$ over some time segments. In addition, BCGAM, CGAM, and LOESS do not allow for the straight forward inference. When comparing to estimated PM$_{2.5}$ from these methods to the measurements from the TEOM, LOESS was the best performing method with an $R^2$ of 0.81 despite having negative estimates of PM$_{2.5}$ for a substantial period of time. The other approaches had higher $R^2$ ranging from 0.57 to 0.72.

\subsection{Posterior visualization and MCMC performance}

To better illustrate how BNMR works and compare the variable selection and clustering approach of BNMR to the variable selection only approach of BISOREG, we show the basis functions used in one sample in Figure~\ref{fig:BP}. Panel~\ref{subfig:BP} shows the BP basis of order 50 as used in the simulation and data analysis. Panel~\ref{subfig:transBP} shows the transformed BP basis $\mathbf{B}\mathbf{A}^{-1}$ as described in Section~\ref{sub:model}. We estimated the posterior mode subset of basis functions used with BNMR and BISOREG. Panel~\ref{subfig:bisoregBP} shows the posterior mode subset of basis functions included into the model with BISOREG.  This is a subset of the full basis expansion shown in Panel~\ref{subfig:transBP}.  Panel~\ref{subfig:bnmrBP} shows the posterior mode combination of basis functions used by BNMR. This includes and intercept and three additional combination basis functions. Each combination basis function is a cluster of one to three of the original transformed basis function in Panel~\ref{subfig:transBP}. BISOREG uses an intercept and nine additional basis functions.  As a result only three unique non-zero slope parameters are estimated with BNMR compared to the nine used by BISOREG. The posterior mode basis functions are shown for all 12 samples in supplemental material. BNMR uses between three and five  combination basis function at its posterior mode with each combination basis function being a cluster of one to five of the original basis functions.

Finally, we compare the MCMC performance of BNMR and BISOREG which both use the same BP basis but have different priors and MCMC approaches. Supplemental Table 5 shows the mean effective sample size and autocorrelation in the posterior sample of $f$. BNMR had a larger average effective sample size than BISOREG (1164 verse 1066 from a posterior sample of 5000 after thinning by 10 from an original sample of 50000) and had lower autocorrelation at lag 1 (0.273 verse 0.375). In part, this efficiency gain can be attributed to the clustering which results in a smaller number of unique regression coefficients being sampled from a truncated multivariate normal distribution. However, there are numerous other differences in the priors and algorithms that likely also contribute to differences in efficiency.

\section{Discussion}

We propose BNMR to estimate a smooth monotone regression function. Our method is motivated by data generated from the MARS aerosol monitor.  This affordable monitor measures the pressure drop across a filter. As particles accumulate on the filter the pressure drop increases. The time-resolved PM$_{2.5}$ concentration is inferred from the first derivative of pressure drop scaled by the total mass collected on the filter. Hence, our objective is to estimate a smooth monotone function and make inference on the scaled derivative of that function.

Our proposed approach uses a BP expansion with a Dirichlet process prior that performs both variable selection and clustering on the regression coefficients for the basis expansion. This formulation enables flexible monotone regression while allowing the model to be null when there is no association between predictor and outcome and allowing the function to be linear when there is no evidence of nonlinearity. Further, we can make coherent, closed-form inference on not only the function being estimated but the derivatives of that function and the scaled derivative of the function. 

Our simulation study showed that BNMR performs similarly to other approaches for smooth nonlinear functions but offers improved inference at smaller sample sizes and when the true function is linear. By both clustering and selecting basis functions, BNMR is self-tuning and results in a smaller parameter space than methods that use variable selection alone.

Our proposed method builds on a substantial body of research on statistical methods to measure or estimate exposure to PM$_{2.5}$, PM components, other environmental pollutants. This includes methods to infer exposures from existing monitoring networks, deployment of networks of portable devices, smartphones, and personal monitors \citep{Calder2008,Rundel2015,Das2017,Huang2018,Finazzi2019}.

\section*{Supplemental Material}

The supplemental material includes replicates of Figures~\ref{fig:BP}, \ref{fig:analysis1}, and \ref{fig:analysis2}  for all 12 runs. It also includes additional simulation results and information on computation time and efficiency. The methods can be implemented with the {\tt R} package {\tt bnmr} available at 
\href{github.com/AnderWilson/bnmr}{github.com/AnderWilson/bnmr}. Data available on request from the authors.

\section*{Acknowledgement}

This work was supported by the U.S. National Aeronautics and Space Administration and the Robert Wood Johnson Foundation through the Earth and Space Air Prize and by the U.S. Centers for Disease Control, National Institute for Occupational Safety and Health (OH010662 and OH011598). 

This work utilized the RMACC Summit supercomputer, which is supported by the National Science Foundation (awards ACI-1532235 and ACI-1532236), the University of Colorado Boulder and Colorado State University. The RMACC Summit supercomputer is a joint effort of the University of Colorado Boulder and Colorado State University.

\bibliography{references}

\end{document}